\documentclass[twocolumn,prb,showpacs,aps,floatfix]{revtex4}
\usepackage{graphicx}
\usepackage{dcolumn}
\usepackage{amsmath}
\usepackage{amssymb}
\usepackage{mathrsfs}
\usepackage{bm}

\begin{document}

\title{Gap maps and intrinsic diffraction losses in 
       one-dimensional photonic crystal slabs}

\author{Dario Gerace}
\author{Lucio Claudio Andreani}

\affiliation{Istituto Nazionale per la Fisica della Materia and
Dipartimento di Fisica ``Alessandro Volta,''\\ 
Universit\`a di Pavia, Via Bassi 6, I-27100 Pavia, Italy}

\date{\today}

\begin{abstract}
A theoretical study of photonic bands for one-dimensional (1D)
lattices embedded in planar waveguides with strong refractive
index contrast is presented.
The approach relies on expanding the electromagnetic field 
on the basis of guided modes of an effective waveguide,
and on treating the coupling to radiative modes 
by perturbation theory.
Photonic mode dispersion, gap maps, and intrinsic diffraction
losses of quasi-guided modes are calculated for the case
of self-standing membranes as well as for Silicon-on-Insulator 
structures.
Photonic band gaps in a waveguide are found to depend 
strongly on the core thickness and on polarization, so that the 
gaps for transverse electric and transverse magnetic
modes most often do not overlap.
Radiative losses of quasi-guided modes above the light line 
depend in a nontrivial way on structure parameters,
mode index and wavevector.
The results of this study may be useful for the design of 
integrated 1D photonic structures with low radiative losses.
\end{abstract}


\maketitle

\section{Introduction}\label{intro}

Photonic crystals embedded in planar dielectric waveguides, 
also known as photonic crystal slabs, 
are intensively investigated as a promising route 
for the tailoring of photonic 
states.$^{1-40}$
Indeed, propagation of light can be controlled in these systems
by the dielectric discontinuity of the slab waveguide
in the vertical ($z$) direction and by the photonic
pattern in the $xy$ plane.
The geometry of a patterned waveguide gives 
considerable freedom in designing photonic structures
(periodic or containing defects) that can be realized
at near-infrared or optical wavelength by lithography and etching.

Most experimental investigations of photonic crystal slabs
with a two-dimensional (2D) or one-dimensional (1D) periodic lattice
concern in-plane transmission\cite{krauss96,labilloy97,benisty99,chow00}
or surface reflectance/transmittance 
measurements\cite{fujita98,astratov99,pacradouni00,cowan01,galli02,
bristow02,patrini02} with the purpose of determining the photonic 
gaps and the band dispersion.
Structures containing defect states 
like linear waveguides in 2D lattices\cite{loncar02}
or microcavities in 1D systems\cite{peyrade02_apl,peyrade02_mne,bristow03}
are also being investigated.
On the theoretical side, the study of photonic crystal slabs 
has been undertaken with plane-wave 
expansion,\cite{fan97,johnson99,sondergaard00,fan02,qiu02}
scattering-matrix methods,\cite{whittaker99,whittaker02,tikhodeev02}
finite-difference time-domain (FDTD) 
calculations,\cite{chutinan00,ochiai01a,hadley02,fan02}
modal methods,\cite{silvestre00,ctyroky01,lalanne01,palamaru01}
and perturbative approaches.\cite{benisty00,ochiai01b,koshino03}
Recently, a finite-basis expansion method 
has been introduced.$^{36-38}$
Most of these papers concern 2D structures, either periodic
or with linear defects.
The theoretical study of 1D structures is restricted to a few 
papers and mostly focused onto the optical response
in both in-plane\cite{ctyroky01,palamaru01}
and out-of-plane\cite{tikhodeev02,bristow02,patrini02,bristow03,koshino03} 
configurations.

Electromagnetic eigenmodes in photonic crystal (PC) slabs
with a periodic pattern have notable differences as compared 
to the ideal reference systems (i.e., not waveguide-embedded),
which are well known from the literature for the cases 
of both 1D\cite{yariv,joannopoulos_book} 
and 2D\cite{joannopoulos_book} periodicities.
A most important issue is the \textit{light-line problem}:
only photonic modes which lie below the light line of the
cladding material (or materials, if the waveguide is asymmetric)
are truly guided and stationary, while those lying
above the light line in the first Brillouin zone are coupled 
to leaky waveguide modes and are subject to intrinsic 
radiative losses.
These \textit{quasi-guided modes} are actually resonances
in a region of continuous energy spectrum, and for this reason
they are more difficult to calculate than truly guided modes
below the light line.
Indeed, while the dispersion of guided modes 
can be obtained by a plane-wave expansion 
with a supercell in the vertical direction\cite{johnson99}, 
the frequencies and especially the losses of quasi-guided modes 
are most commonly studied by FDTD calculations.\cite{sakoda_book}
Another important feature of photonic crystal slabs
is the blue shift of the eigenmodes due to vertical confinement
in the planar waveguide. This effect, which is more pronounced
for slabs with strong out-of-plane refractive index contrast,
implies that the frequencies of photonic bands and gaps
depend on all parameters of the planar waveguide 
(layer thicknesses and refractive indices) and can differ
substantially from those of the reference 1D or 2D system.
Finally, the eigenmodes of photonic crystal slabs 
can be put in one-to-one correspondence with those 
of the reference system only when the frequency 
is sufficiently low for the waveguide to be monomode.
The cut-off frequency of second- and higher-order modes
also depends on slab parameters and on the photonic lattice.

\begin{figure}[t]
\begin{center}
\includegraphics[width=0.3\textwidth]{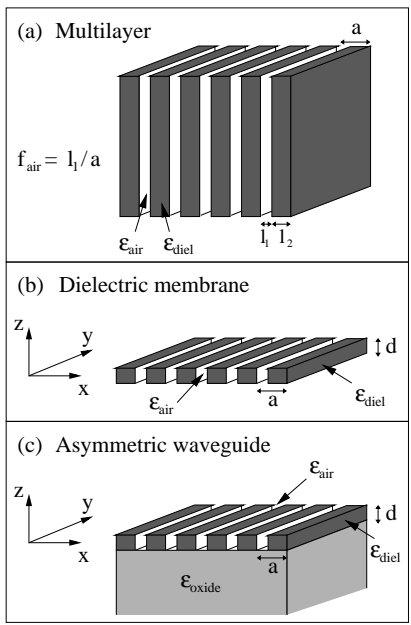}

\caption{\label{structure}
         Photonic structures studied in this
         work: (a) Ideal one-dimensional photonic crystal,
         with period $a$ and air fraction
         $f_{\mathrm{air}}=l_1/a$. (b) Photonic crystal slab
         consisting of a self-standing, patterned dielectric
         core (\textit{air bridge} or
         \textit{membrane}) of thickness $d$ surrounded by air.
         (c) Photonic crystal slab, with the pattern defined in a
         high-index dielectric core of thickness $d$ sandwiched
         between air and an insulating oxide substrate.
         Throughout this work we assume:
         $\epsilon_{\mathrm{diel}}=12$,
         $\epsilon_{\mathrm{oxide}}=2.1$,
         $\epsilon_{\mathrm{air}}=1$.
}
\end{center}
\end{figure}

\begin{figure}[b]
\begin{center}
\includegraphics[width=0.5\textwidth]{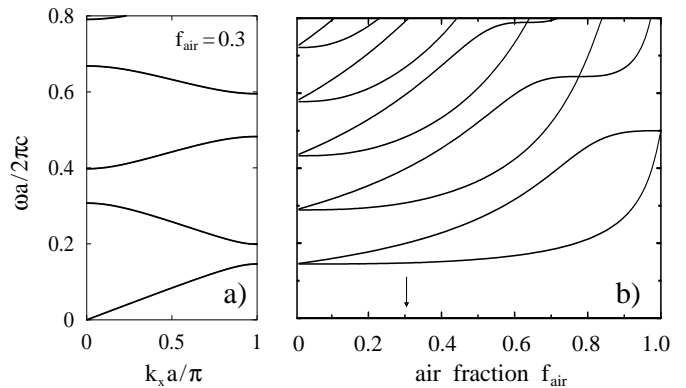}

\caption{\label{ideal}
         Ideal multilayer: (a) Photonic
         bands for $f_{\mathrm{air}}=0.3$;
         TE and TM modes are exactly degenerate.
         (b) Gap map, i.e., band edges as a function of
         air fraction; the value of $f_{\mathrm{air}}$
         corresponding to the calculation given in (a)
         is indicated by an arrow.
}

\end{center}
\end{figure}

In this work we present a systematic study of photonic bands,
gap maps and diffraction losses for 1D photonic crystal slabs, 
that is 1D photonic lattices
like those of a distributed Bragg reflector 
(see Fig.~\ref{structure}a for the 1D reference system).
These are defined in two types of
waveguides with strong refractive 
index contrast: the self-standing membrane or \textit{air bridge} 
(Fig.~\ref{structure}b) and the asymmetric photonic crystal slab
in which only the core layer is patterned (Fig.~\ref{structure}c).
The latter structure is typically realized with
the Silicon-on-Insulator (SOI) system 
but may also be realized with GaAs on an oxide layer.
We assume the following values of the dielectric constants:
$\epsilon_{\mathrm{diel}}=12$ for the high-index core layer (as 
appropriate to Si or GaAs below the band gap), 
$\epsilon_{\mathrm{air}}=1$, and $\epsilon_{\mathrm{oxide}}=2.1$ 
(as appropriate for SiO$_2$ or other oxides).
The periodic patterning is taken along the $x$ direction and
throughout this paper we assume $k_y=0$.
The gap maps are calculated as a function of air fraction
in the core layer and for different values of the core thickness,
thereby exploring a wide range of parameters of experimental interest.
The dependence of the radiation losses on frequency, polarization,
and air fraction is also calculated and discussed.

The photonic bands and gap maps of a distributed Bragg reflector
are obviously well known and are exemplified in 
Fig.~\ref{ideal}.\cite{lambda4}
Notice that the photonic bands of Fig.~\ref{ideal}a (which refer to 
an air fraction $f_{\mathrm{air}}=0.3$) as well as the gap map of 
Fig.~\ref{ideal}b are degenerate for transverse electric (TE) 
and transverse magnetic (TM) polarizations with respect
to the plane of incidence: 
this degeneracy is lifted in a waveguide because
the confinement-induced shift is polarization-dependent,
as was already shown experimentally.\cite{patrini02}
One of the goals of the present paper is to establish
whether a complete band gap for both polarizations
can occur in a waveguide-embedded 1D photonic structure.

Related concepts have been studied in the context
of dielectric waveguide gratings, also called
resonant grating filters.$^{41,44-59}$
These kinds of diffraction gratings may support 
guided and leaky modes. The latter are resonantly
coupled to an external light beam and give rise
to narrow resonances in reflection or transmission,
which may be used for polarization-dependent filters\cite{magnusson92}
or for enhanced nonlinear optical effects.\cite{neviere95,neviere_book}
Most of the research concentrated on systems
with a weak dielectric modulation,
e.g., surface relief gratings
for filtering and distributed feedback,\cite{yariv,popov93}
waveguides with a weak refractive index contrast 
within the core region\cite{li92,stancil96,tamir97}
and/or which are modeled by a single Fourier component
of the dielectric function.$^{44,51-53,58}$
For an extensive list of previous literature along these lines
and of the different kinds of theoretical methods used we 
refer to the book by Loewen and Popov.\cite{popov_book}
In all these cases, which can be treated at least qualitatively
by coupled-mode theory, the dispersion of the waveguide mode
is only weakly modified by the dielectric modulation
and photonic bandgap effects are very small.
Specific waveguide grating structures 
with strong refractive index modulation in the plane 
leading to an appreciable photonic gap 
have been studied in Refs.~\onlinecite{brundrett98,brundrett00} 
for the case of TE polarization,
and in Refs.~\onlinecite{lalanne00,cao02}
for both TE and TM polarizations.
In these strongly modulated cases a rigorous coupled-wave analysis
(also called the Fourier modal method) is necessary and has been used.
We point out that the focus of the present work is quite different from
all these papers, in particular for what concerns
the systematic calculation of gap maps and losses as a function 
of frequency and of various structure parameters.

This work is organized as follows. 
In Sec.~II we give a short description of the theoretical method
for calculating photonic bands and intrinsic losses in a waveguide.
In Sec.~III we discuss a few examples of photonic mode dispersion, 
both for symmetric and asymmetric 1D photonic crystal slabs.
Section IV contains detailed results for 1D gap maps 
in a membrane or in an asymmetric waveguide for different 
values of the core thickness.
In Sec.~V we present results for intrinsic losses of 
quasi-guided modes. In Sec.~VI we give a few closing remarks.

\section{Method}\label{method}

The approach adopted here, which was already introduced
in Refs.~\onlinecite{andreani02_ieee,andreani02_pss},
relies on a finite-basis expansion in order to transform 
the second-order equation for the magnetic field in
a linear eigenvalue problem. 
The basis consists of the guided modes of an effective
homogeneous waveguide, where the dielectric constant of each 
layer is defined by the spatial average of the dielectric 
constant $\epsilon(\mathbf{r})$ over the photonic pattern.
The in-plane Bloch vector $\mathbf{k}$ is obviously conserved
modulo a reciprocal lattice vector $\mathbf{G}$.
The off-diagonal components $\epsilon_{\mathbf{G},\mathbf{G}'}$
of the dielectric tensor give rise to band splittings 
and to a folding of the photonic modes in the first Brillouin zone.
Some of the photonic modes may lie below the cladding light line 
and be truly guided, however most (sometimes all) of them
fall above the light line in the first Brillouin zone.
Coupling of these modes to leaky modes of the effective waveguide
is taken into account by time-dependent perturbation theory,
which leads to an expression for the imaginary part of the mode 
frequency in terms of the photonic density of states at fixed in-plane 
wavevector.\cite{ochiai01b,andreani02_pss,comparison}
This procedure is formally analogous 
to Fermi's golden rule in quantum mechanics.
When the waveguide is asymmetric (like in the case
of the SOI structure), care must be taken to express
the leaky modes in terms of outgoing states and to relate
them to the respective state densities.\cite{carniglia71} 

The approximations made in the present approach are as follows 
(for a fuller discussion 
see Refs.~\onlinecite{andreani02_ieee,andreani02_pss}).
The effective dielectric constant of the homogeneous waveguide,
which defines the basis of guided modes for the expansion, 
is chosen to be
\begin{equation}
\epsilon_{\mathrm{eff}}=f_{\mathrm{air}}
\epsilon_{\mathrm{air}}+(1-f_{\mathrm{air}})\epsilon_{\mathrm{diel}}\, ,
\label{media}
\end{equation}
for both TE and TM polarizations. This choice is by no means
unique, although it is the exact definition of the effective
dielectric constant for TE polarization and anyway
when the electric field is perpendicular 
to the direction of periodicity.\cite{agranovich85}
For TM-polarized modes,
which have electric field components along $x$ and $z$,
the situation is more complex.\cite{tmpol}
It is known from the literature that the $x$ component
of the electric field is subject to an effective dielectric constant
that is obtained from the inverse averaging rule,\cite{agranovich85}
therefore a different choice of $\epsilon_{\mathrm{eff}}$ 
in the patterned region could be suggested for TM modes. 
Choices of $\epsilon_{\mathrm{eff}}$ differing from Eq.~(\ref{media})
do not lead to any appreciable change of the results
above the mode cutoff, as we have verified. The frequency position
of the cutoff does depend on the choice of $\epsilon_{\mathrm{eff}}$,
especially for large air fractions, however a comparison with
exact scattering matrix calculations~\cite{patrini02,andreani02_ieee}
shows that the average dielectric constant defined by Eq.~(\ref{media})
gives very good agreement with the frequencies and cutoff positions
obtained from the exact calculations.
It should also be noted that the electromagnetic field close to mode
cutoff is mostly extended in the claddings, where the dielectric
constants are homogeneous for the airbridge and SOI structures
studied in the present paper.

The number of reciprocal lattice vectors $\mathbf{G}$ is limited 
by a finite cut-off, like for usual plane-wave 
calculations,\cite{ho90} and in addition a restricted number 
of guided modes of the effective
waveguide is kept in the expansion. For the calculations shown
in this work, a number of 31 plane waves is usually taken 
in the basis set and is sufficient for convergence
with better than percent accuracy. 
The number of guided modes in the expansion is not found to be 
critical in the energy range considered and is usually taken 
to be $\le 8$.
For the quasi-guided modes, neglect of the second-order shift
due to coupling to leaky modes introduces an error of less than 
a few percent in the photonic frequencies.
All these approximations are justified {\em a posteriori}
by the close agreement of the calculated photonic frequencies
with those obtained from reflectivity calculations\cite{andreani02_ieee}
made with the exact scattering-matrix method.\cite{whittaker99}
Finally, calculating the radiative losses of quasi-guided modes
by first-order perturbation theory is justified by the fact
that the imaginary part of the frequency is much smaller
than the real part, as shown by the results below.

\section{Photonic bands}\label{bands}

\begin{figure*}[t]
\begin{center}
\includegraphics[width=\textwidth]{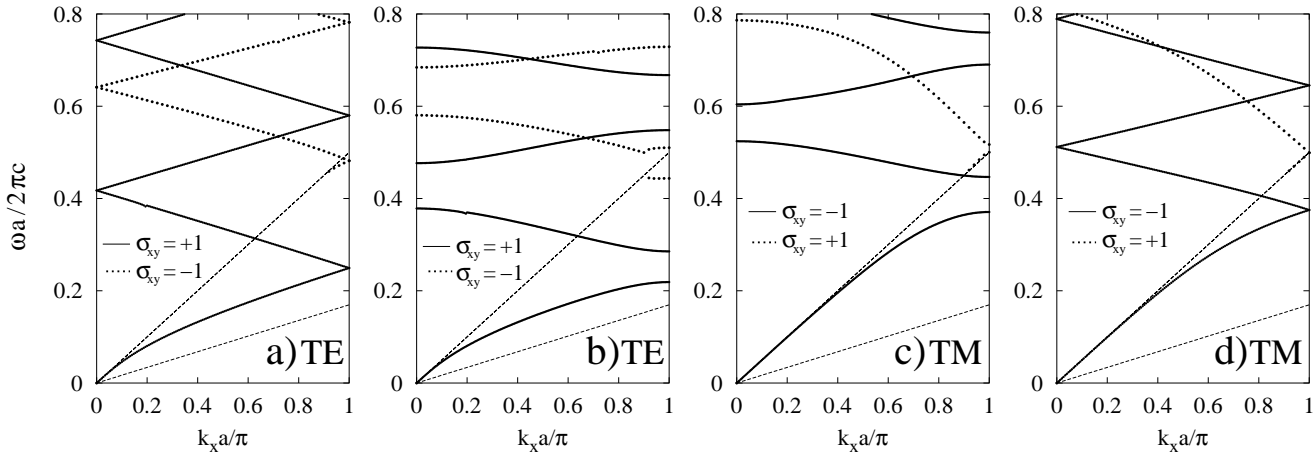}

\caption{\label{bandair}
         Photonic bands for the membrane structure of
         Fig.~\ref{structure}b. The dashed lines represent
         the dispersions of light in air
         and in the average core layer.
         (a) TE, (d) TM dispersion curves,
         folded in the first Brillouin zone,
         for a uniform dielectric membrane with
         $\epsilon_{\mathrm{eff}}=8.7$ and thickness $d/a=0.4$;
         (b) TE, (c) TM photonic bands
         for the patterned structure with
         $f_{\mathrm{air}}=0.3$, $d/a=0.4$.
}

\end{center}
\end{figure*}

The photonic bands of the strong confinement symmetric slab,
corresponding to the
system schematically shown in Fig.~\ref{structure}b, 
are displayed in Fig.~\ref{bandair}b and \ref{bandair}c for 
a core thickness $d=0.4a$ and an air fraction $f_{\textrm{air}}=0.3$.
The bands are plotted by using dimensionless frequency 
$\omega a/(2\pi c)=a/\lambda$ as a function of in-plane 
wave vector $k_x a/\pi$ in the first Brillouin zone.
The photonic dispersion curves of the patterned structure 
are compared to those of a uniform dielectric slab suspended in air 
(Figs.~\ref{bandair}a and \ref{bandair}d) 
with a spatially averaged dielectric constant given by Eq. (\ref{media}),
that is $\epsilon_{\mathrm{eff}}=8.7$ in the present case. 
The guided modes of Figs.~\ref{bandair}a and \ref{bandair}d
represent the basis set for the expansion method discussed in
the previous Section.
The dispersion for the average dielectric slab is presented 
in the reduced zone scheme, allowing for a direct comparison 
with the corresponding photonic bands of the patterned waveguide.
We have classified the guided modes according to mirror symmetry 
with respect to the plane of incidence $\mathbf{k}z\equiv xz$ 
(we use $\sigma_{xz}$ to denote this operation) 
and with respect to the $xy$ plane ($\sigma_{xy}$ operation). 
The modes whose electric field component
lies in the $xy$ plane are referred to as TE, 
and are odd with respect to specular reflection 
through the plane of incidence ($\sigma_{xz}=-1$); 
the modes whose magnetic field lies in the $xy$ plane are 
labelled as TM and are even with respect to mirror plane $xz$
($\sigma_{xz}=+1$).\cite{ochiai01a} 
These modes can be classified further as even ($\sigma_{xy}=+1$) 
or odd ($\sigma_{xy}=-1$) with respect to specular reflection 
through the $xy$ plane, thus giving four different types of 
guided eigenfunctions for the electromagnetic field. 
We can separately compare Fig.~\ref{bandair}a to \ref{bandair}b 
and Fig.~\ref{bandair}c to \ref{bandair}d. 
It is clearly seen that for both TE and TM modes the 
periodic patterning of the dielectric slab introduces band gaps 
around the degenerate points of the average slab dispersion 
curves ($k_x=0$ and $k_x=\pm\pi/a$), owing to the off-diagonal
components of the inverse dielectric tensor. 
There is one-to-one correspondence between average
slab and 1D PC slab modes. 
The first-order modes (TE even and TM odd) have no cut-off
frequency, as is well known for a symmetric waveguide.
The second-order guided modes have a finite cut-off frequency,
which is degenerate for TE and TM modes.
The second-order mode, represented by dotted lines,
has $\sigma_{xy}=-1$ for TE polarization, 
while it has $\sigma_{xy}=+1$ for TM polarization.

\begin{figure}[b]
\begin{center}
\includegraphics[width=0.5\textwidth]{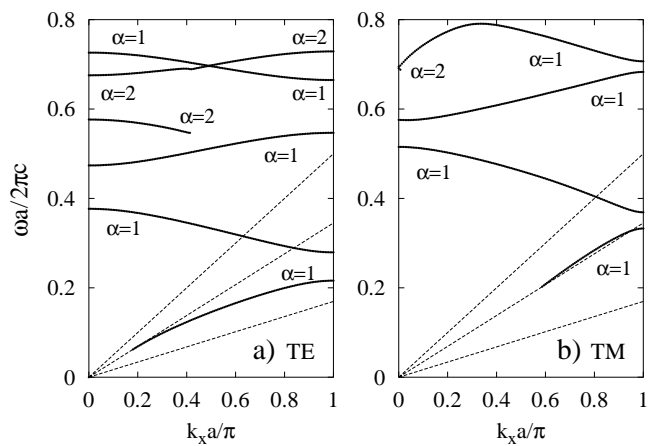}

\caption{\label{bandsoi}
         Photonic bands for the patterned dielectric-on-insulator
         structure of Fig.~\ref{structure}c
         with $f_{\mathrm{air}}=0.3$, $d/a=0.4$.
         The dashed lines are the light dispersions in
         the effective core and in the upper and lower claddings;
         $\alpha$ labels the order of the guided mode.
         (a) TE, (b) TM modes.
}

\end{center}
\end{figure}

A second point should be remarked by comparing the photonic bands
of Figs.~\ref{bandair}b and \ref{bandair}c to the bands of an ideal
multilayer. 
The first photonic band gap appears between 0.15 and 0.20 in the 
ideal 1D case, and between 0.22 and 0.28 for the lowest TE 
mode in the PC slab, due to
the confinement effect along the vertical ($z$) direction. 
The gap between the first and the second band opens between 0.37 
and 0.45 when considering TM modes: these values are strongly 
blue shifted with respect to both the multilayer and the 
TE modes of the PC slab. 
Thus the confinement effect manifests itself in the 
blue shift of the eigenfrequencies of the electromagnetic field
with respect to the ideal multilayer,
and moreover in the removal of degeneracy between TE and TM modes:
the latter effect is due to the stronger confinement of TM 
compared to TE modes in the planar waveguide.\cite{yariv}  
We also notice that all the band gaps, except for the first one, 
lie in the region of guided resonances, and could
be experimentally tested by external reflectance measurements.  
The first band gap, either TE even or TM odd, 
is instead in the region of truly guided modes. A complete band gap
common to both polarizations can also be seen around 
$\omega a/(2\pi c)\sim 0.4$, where the second-order TE gap 
overlaps the first-order TM gap.
As we will see in the next section, this is rather a coincidence
for 1D PC slabs.

In Fig.~\ref{bandsoi} we display the photonic bands 
for the asymmetric structure 
represented in Fig.~\ref{structure}c. The dashed lines
are the dispersions of photons in air, substrate and effective 
dielectric core.  
The parameters used in these calculations are $d/a=0.4$, 
$f_{\textrm{air}}=0.3$, allowing for a direct comparison 
with the results of Figs.~\ref{bandair}b and \ref{bandair}c.
Owing to the asymmetry of the vertical waveguide, $\sigma_{xy}$ 
is no more a symmetry operation: the modes can only be classified 
as odd  (TE, Fig.~\ref{bandsoi}a) or even (TM, Fig.~\ref{bandsoi}b) 
with respect to the plane of incidence. 
However, we have indicated the approximate order of the 
waveguide mode by the index $\alpha$ in Fig.~\ref{bandsoi} 
(this can be defined only when the modes are well separated 
in frequency, otherwise mixing and anticrossing effects occur).
For an asymmetric slab there are no modes starting at 
$\omega=0$.\cite{yariv} 
By comparing Figs.~\ref{bandair} and \ref{bandsoi}, 
we notice that the lowest TE mode of the asymmetric 1D PC slab
is in quantitative agreement with the first-order TE mode of the
PC membrane; instead, the TM modes of the asymmetric slab 
are somewhat less confined than those of the PC membrane.
It is important to stress that the modes lying between 
the two claddings light lines (oxide and air in this case) 
are evanescent in air, but leaky in the substrate. 
These modes have finite radiative losses, 
as we will see in section~\ref{losses}.   
We also notice that no complete band gap is present in 
the asymmetric 1D PC slab, at variance with the corresponding 
symmetric structure. 
The results shown in Fig.~\ref{bandsoi} are conceptually similar
to Brillouin diagrams calculated for TE polarization
in the case of an asymmetric waveguide grating structure.\cite{brundrett00}
We point out that the described features of photonic band
structures for an asymmetric 1D PC slab were 
experimentally verified by variable angle surface reflectance
performed on SOI structures, for both TE and TM modes.\cite{patrini02} 

\section{Gap maps}\label{maps}

In this section we present a complete set of gap maps
for waveguide-embedded 1D photonic crystals.
We consider a \textit{band gap} 
as a frequency region where no photonic modes exist, 
either truly guided or quasi-guided above the light line.
We present the gap maps for modes with TE or TM polarizations,
i.e., with definite parity with respect to the
vertical mirror symmetry $\sigma_{xz}$: 
this convention applies to symmetric as well as asymmetric
vertical waveguide structures.
For the case of the asymmetric structure, 
for which the lowest-order waveguide mode has a finite cut-off,
only the frequency region above the lowest-order cut-off is 
physically relevant.

\subsection{Dielectric membrane}\label{air}

In Fig.~\ref{mapair} we display the calculated gap maps 
for the air bridge structure
of Fig.~\ref{structure}b. We show the maps for three different
slab thicknesses, namely $d/a=0.2$, $0.4$, and $0.8$.
We display in black the true complete band gap, 
i.e., the frequency region in which 
no photonic modes (or resonances) are allowed for any polarization.
The gap maps are shown for air fraction varying from 0 to 0.7, 
which represent a realistic range for practical realization.
The solid lines in Fig.~\ref{mapair} represent the cut-off 
frequency of the second-order waveguide mode, 
which is given by   
\begin{equation}
\begin{small}
\frac{\omega_{\mathrm{c}}a}{2\pi c}=
\frac{a}{ 2d\sqrt{ \epsilon_{\mathrm{eff}}-\epsilon_{\mathrm{air}}}}
\label{sim}
\end{small}
\end{equation}
and is the same for both polarizations.

An important feature that we can see from Fig.~\ref{mapair} 
is that for $d/a=0.2$, the TE gap map is qualitatively similar
to the ideal multilayer one (see Fig.~\ref{ideal}b)
with a blue shift arising from the confinement effect.
The band gaps for TM modes are shifted to much higher frequencies
and some complete band gaps start to appear only at 
$a/\lambda\sim 0.7$. The gap map is more complex for $d/a=0.4$, 
due to the appearence of higher-order waveguide modes at low frequency. 
Nevertheless, for $a/\lambda\lesssim 0.4$ the slab is still monomode, 
and a large complete band gap opens in a wide range of 
air fractions (Fig.~\ref{mapair}b). 
This complete gap comes from the overlap
of the first TM gap (at the Brillouin zone edge, see 
Fig.~\ref{bandair}c) and the second TE one (at the zone 
center, Fig.~\ref{bandair}b). No complete band gap has been 
found for other values of $d/a$ (calculations not shown).   
For $d/a=0.8$ the photonic band structure is quite complex because
the slab becomes multimode already at low frequencies.
The band gap in TM modes is still present around $a/\lambda\sim 0.4$, 
but no complete band gap exists
because of the presence of second-order TE modes.
The conclusions from these results are the following:
(i) the TE gap map in a waveguide resembles the ideal 1D one
only below the cut-off of second-order modes,
(ii) the TM gap map is very sensitive to the structure parameters,
and (iii) a complete gap for both polarizations
is calculated to occur only for a core thickness around $d/a=0.4$.

\subsection{Asymmetric waveguide}\label{soi}

\begin{figure}[t]
\begin{center}
\includegraphics[width=0.35\textwidth]{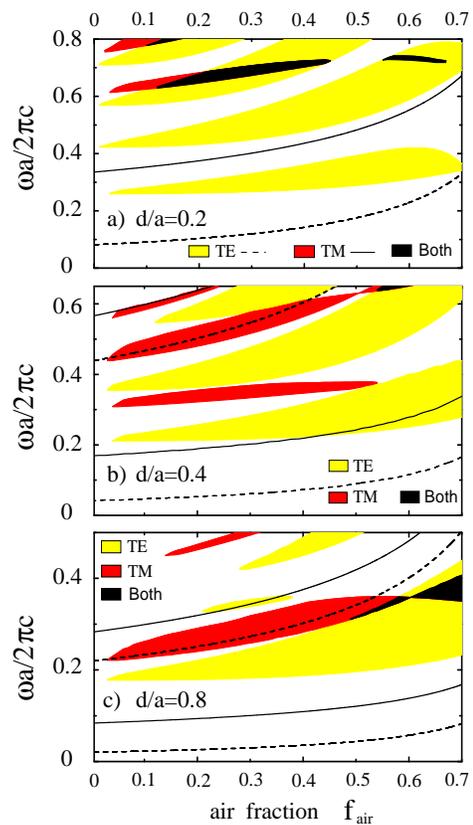}

\caption{\label{mapsoi}
         Gap maps for the asymmetric PC slab structure
         of Fig.~\ref{structure}c.
         Dashed (solid) lines represent the cut-off frequencies
         of the first- and second-order waveguide modes
         for TE (TM) polarization.
         (a) Core thickness $d/a=0.2$, (b) $d/a=0.4$, (c) $d/a=0.8$.
}

\end{center}
\end{figure}

In Fig.~\ref{mapsoi} we show the calculated gap maps of the asymmetric 
1D PC slab of Fig.~\ref{structure}c, for core thicknesses $d/a=0.2$, 
$0.4$, and $0.8$. 
The gap maps of Fig.~\ref{mapsoi} show notable differences
as compared to those of the PC membrane. 

One of the peculiarities of the asymmetric structure
is the existence of a finite cut-off frequency 
for the lowest-order TE and TM modes.
The cut-off frequency as a function of air fraction is plotted 
with dashed lines for TE modes, and with solid lines for TM modes. 
The values for the cut-off frequencies obtained by the present approach
coincide with those following from the expression for a
uniform, asymmetric slab. The formula is\cite{yariv}
\begin{equation}
\begin{small}
\frac{\omega_{\mathrm{c}}a}{2\pi c}=
      \frac{a}{2d\sqrt{\epsilon_{\mathrm{eff}}-\epsilon_{\mathrm{oxide}}}}
       \left[m+\frac{1}{\pi}\textrm{arctan}
       \left(r\frac{\sqrt{\epsilon_{\mathrm{oxide}}
                                      -\epsilon_{\mathrm{air}  }}}
                                {\sqrt{\epsilon_{\mathrm{eff}} 
                                -\epsilon_{\mathrm{oxide}}}}
    \right)\right] \label{asim}
\end{small}
\end{equation}
where $r=1$ for TE modes, $r=
\epsilon_{\mathrm{eff}}/\epsilon_{\mathrm{air}}$ for TM modes, 
and $m\ge0$ is an integer.

\begin{figure}[t]
\begin{center}
\includegraphics[width=0.35\textwidth]{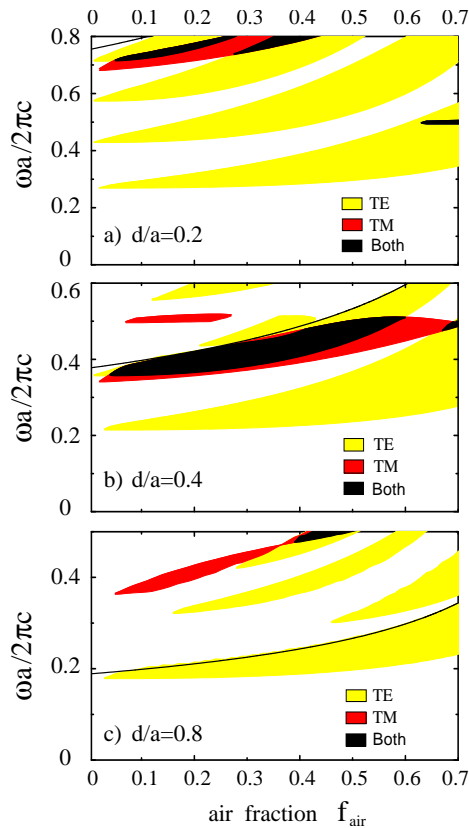}

\caption{\label{mapair}
         Gap maps for the membrane structure of Fig.~\ref{structure}b,
         as a function of the air fraction $f_{\mathrm{air}}=l_1/a$.
         Solid lines represent the cut-off frequency
         of the second-order waveguide mode.
         (a) Core thickness $d/a=0.2$, (b) $d/a=0.4$, (c) $d/a=0.8$.
}

\end{center}
\end{figure}

For $d/a=0.2$ the asymmetric 1D PC slab has only first-order
TE and TM modes in the whole frequency range shown.
The TE band gaps are again qualitatively similar to those
of the ideal 1D multilayer, with a confinement effect
which is close to that of the membrane case (Fig.~\ref{mapair}a);
the TM gaps are instead shifted to much higher frequencies
as compared to the 1D multilayer.
The first TE band gap is in a region below the cut-off
of the first-order TM mode, thus it may be considered
as a complete band gap.
For $d/a=0.4$ and 0.8 a second-order TE cut-off appears 
at frequencies around 0.44 and 0.22, respectively:
the TE gap map is similar to that of the 1D multilayer
only below the second-order cut-off frequency.
The TM gaps are always very different from those
of the ideal 1D case and also quite different
from those of the PC membrane: TM modes are seen
to be extremely sensitive to the structure parameters
(core thickness and claddings dielectric constants).
As it can be seen by comparing Figs.~\ref{mapair}b and \ref{mapsoi}b, 
the complete band gap for $d/a=0.4$ occurs for the
particular case of a 1D PC membrane but not in the
asymmetric PC slab. 
For $d/a=0.8$ a complete band gap resulting from the overlap
of the first TE and TM gaps appears around $a/\lambda\sim 0.3$ for 
$f_{\mathrm{air}}\gtrsim 0.5$.

As a general remark, the numerical results for photonic bands and gaps
previously shown relate only to the real part of the frequency
and do not consider the effect of coupling to radiative waveguide modes.
Thus the physical relevance of a photonic band dispersion 
is expected to decrease on increasing the frequency far from the light line.
We also notice that the concept of mode cut-off for resonant modes
is not clearly defined when radiative broadening is taken into account.
For these reasons, the gap maps calculated here are expected
to be more useful in the low frequency region,
in particular for the band gaps which open 
below the second-order cut-off lines
shown in Figs.~\ref{mapair} and \ref{mapsoi}.

\section{Diffraction losses}\label{losses}

\begin{figure}[t]
\begin{center}
\includegraphics[width=0.5\textwidth]{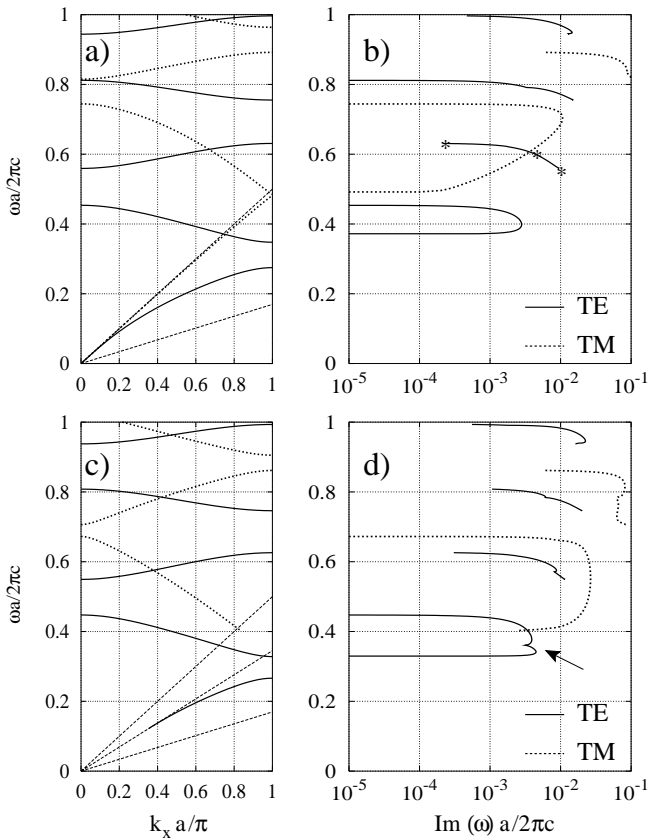}

\caption{\label{loss}
   (a) Photonic bands and (b) imaginary part of frequency
   for a symmetric 1D PC membrane;
   (c) photonic bands and (d) imaginary part of frequency
   for an asymmetric 1D PC slab.
   Parameters are $d/a=0.2$, $f_{\mathrm{air}}=0.3$.
   Solid (dashed) lines are TE (TM) modes.
   The three markers in (b) correspond to the points marked in
   Fig.~\ref{loss2} (see text).
   The arrow in (d) denotes the cusp, which corresponds to the
   second TE band in (c) crossing the air light line.
}

\end{center}
\end{figure}

To complete our analysis of 1D PC slabs we have to address also the
imaginary part of frequency, which gives information about
the radiative losses due to out-of-plane diffraction.
This is done by using time-dependent perturbation theory 
for the electromagnetic problem,
as previously discussed in Section~\ref{method}. 
We display in Fig.~\ref{loss} the results for parameters $d/a=0.2$ 
and $f_{\mathrm{air}}=0.3$, for both a PC membrane 
and an asymmetric PC slab.
In Figs.~\ref{loss}a and \ref{loss}b the band diagram and 
the corresponding imaginary part of frequencies are shown for the 
symmetric 1D PC slab. 
In Fig.~\ref{loss}a the waveguide 
is monomode for both TE and TM polarizations.
In Fig.~\ref{loss}b we show the dimensionless quantity 
Im$(\omega)a/(2\pi c)$, corresponding to each photonic band 
of Fig.~\ref{loss}a, as a function of mode frequency.
The imaginary part is generally much smaller than the real part,
indicating the validity of the perturbative treatment adopted.
The losses go to zero when the mode crosses the light line in air
and becomes truly guided.
It is clear from the figure that the radiative losses 
generally increase on increasing the photonic band index, 
however the behavior of the losses within a given photonic 
band is nontrivial and has to be studied in each specific case.
The guided resonances at the 
Brillouin zone center present a quite interesting behavior.
In fact, the second and the fourth TE bands have zero linewidth at 
$k_x=0$ (their frequencies are $\omega a/(2\pi c)=0.45$ and 0.81, 
respectively) while the third and the fifth band have finite 
radiative widths at $k_x=0$ ($\omega a/(2\pi c)=0.56$ and 0.94).
A similar behavior holds also for TM modes.
These numerical results could be probed by variable angle surface
reflectance experiments made on 1D PC membranes:
the imaginary part of the frequency can be extracted
from the linewidth of spectral structures in reflectance
that correspond to photonic modes.\cite{pacradouni00,cowan01}

\begin{figure}[b]
\begin{center}
\includegraphics[width=0.5\textwidth]{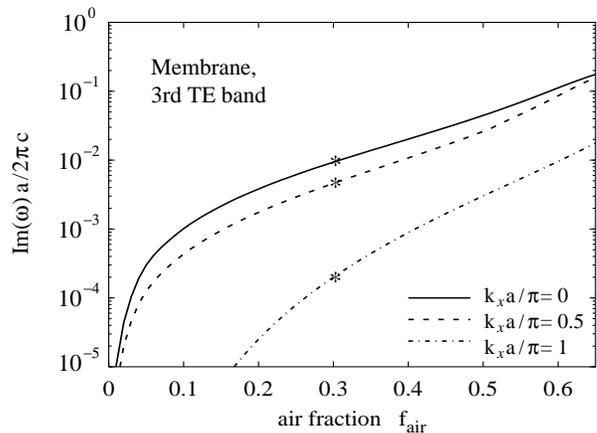}

\caption{\label{loss2}
         Imaginary part of photonic frequencies as a
         function of the air fraction for the PC membrane
         of thickness $d/a~=~0.2$; the three curves correspond to
         different wavevectors ($k_x~=~0$, $k_x~=~\pi/2a$,
         $k_x~=~\pi/a$) in the first Brillouin zone of the third
         TE photonic band.
         The three points marked on the curves correspond to the three
         markers of Fig.~\ref{loss}b, where $f_{air}~=~0.3$.
}

\end{center}
\end{figure}

In Figs.~\ref{loss}c and \ref{loss}d we show the results for 
an asymmetric 1D PC slab with the same thickness and air 
fraction.\cite{brillouin}
The radiative losses shown in Fig.~\ref{loss}d
display quite the same behavior as in the membrane case.
They are about two times larger than the corresponding losses 
of Fig.~\ref{loss}b: this is due to the asymmetry of the 
vertical waveguide, which implies that a quasi-guided mode above 
the light line is coupled to all radiative modes of the 
effective waveguide at the same frequency, without the parity 
selection rule which holds instead in the symmetric case.
A similar behavior was found in the calculation of spectral
properties of deep 1D gratings.\cite{ctyroky01}
Moreover, the higher-order modes (either TE or TM) now have
a finite Im($\omega$) also at $k_x=0$:
this is due to the additional diffraction channels for radiative 
losses which are present in the asymmetric waveguide.
Moreover, we notice that the modes whose frequencies lie between 
the light lines of air and oxide claddings are not truly guided, 
i.e., they are evanescent in air but leaky in the substrate. 
Thus, the crossing between a band and the light line in air does 
not cause the linewidth of the photonic resonance to go to zero:
rather, Im($\omega$) has a cusp 
(marked by an arrow in Fig.~\ref{loss}d)
when the light line in air is crossed.
Similar features can by recognized in Fig.~\ref{loss}d
at higher frequencies: they arise whenever a photonic mode crosses
a cladding light line folded in the first Brillouin zone.
These notable features of Im($\omega$) are not a numerical artifact,
but rather they correspond physically to the opening or closing
of diffraction channels for radiative losses.

We also found that the imaginary part of frequency
increases on increasing the air fraction in the investigated range,
as shown in Fig.~\ref{loss2}, where Im($\omega a/2\pi c$) is plotted 
for the PC membrane of thickness $d/a=0.2$. 
A similar behavior is found also for the 
asymmetric structure (not shown here) and an increase
of the photonic mode linewidth with the air fraction
was already stated experimentally.\cite{kawai01,patrini02} 
The three curves of Fig.~\ref{loss2} correspond to the evolution
of the losses for the third TE photonic band 
at three different points in the first Brillouin zone.
When $f_{air}=0.3$ the corresponding photonic band has frequencies around
$\omega a/(2\pi c)=0.6$ (see Fig.~\ref{loss}a), and the three points
marked in Fig.~\ref{loss2} correspond to those marked in Fig.~\ref{loss}b. 
Notice that the losses vary in a logarithmic scale
and become extremely small either towards the homogeneous waveguide
limit at low air fraction or close to the Brillouin zone edge.
While the filling fraction dependence of the losses
is similar for all bands and polarizations, 
the wavevector dependence changes from band to band,
as it appears from Figs.~\ref{loss}b,d.
It can be concluded that for the present waveguide-embedded
1D photonic structures, the spectral linewidth
of quasi-guided photonic modes can vary by several
orders of magnitude and it depends in a nontrivial way 
on the structure parameters
as well as on the angle of incidence, mode index, and polarization.

\section{Conclusions}

We have studied the photonic bands and the gap maps of 
one-dimensional photonic crystal slabs made of a high 
refractive index material sandwiched between low index claddings, 
thus providing strong confinement of electromagnetic waves along 
the vertical direction. 
The method adopted yields the frequencies of photonic modes both
below and above the light line, and therefore allows treating
guided and quasi-guided modes on the same footing.

Photonic modes in a 1D photonic crystal waveguide can be put 
in one-to-one correspondence with those of the ideal 1D 
reference system only when the waveguide is monomode. 
Under this condition, the TE gap maps are qualitatively similar
to the ideal 1D ones considering the blue shift in the waveguide.
This confinement effect is considerably more pronounced for
TM-polarized than for TE-polarized modes, thereby leading
to a polarization splitting which depends in a sensitive way
on the structure parameters. 
As a consequence, a complete band gap common to TE and TM
polarizations is generally not found in 1D photonic crystal 
slabs, except for special values of the parameters.

The radiative losses of guided resonances due to out-of-plane
diffraction depend in a sensitive way on waveguide
parameters, mode index, and frequency. 
For some modes the imaginary part
of the frequency vanishes at the Brillouin zone center,
thus even above the light line it is possible to find
photonic modes with very low losses.
In general, the losses are predicted to be higher for 
Silicon-on-Insulator structures as compared to photonic 
crystal membranes, due to the asymmetry of the planar 
waveguide. All these results are related to the presence
of diffraction channels for radiative couplings.

The quantitative results for the complex frequency dispersion 
of quasi-guided modes may be experimentally tested 
by performing reflectance or transmittance measurements 
with light incident on the PC slab surface. Modes below the
light line can be probed in waveguide transmission experiments.
Moreover, the results of the present work may be useful for 
designing 1D PC slabs with gaps at specified frequencies, 
regions of monomode propagation, and low diffraction losses.
They might also be used for the design of advanced 1D
structures like filters and microcavities.
The same theoretical approach can be applied to PC slabs
with various 2D patterns.
 
\begin{acknowledgments}
The authors are grateful to M.~Agio, M.~Galli
and M.~Patrini for several helpful discussions.
This work was supported by MIUR through Cofin and FIRB programs
and by INFM through PRA PHOTONIC.
\end{acknowledgments}


\begin{thebibliography}{99}

\bibitem{ieee} For recent reviews, see e.g.\ papers
in IEEE J.\ Quantum Electron.\ $\mathbf{38}$, 
Feature Section on Photonic Crystal Structures
and Applications, edited by T.F.~Krauss and T.~Baba,
pp. 724-963 (2002).

\bibitem{krauss96} 
T.F.~Krauss, R.M. De~La~Rue, and S.~Brand, 
Nature $\mathbf{383}$, 699 (1996).

\bibitem{fan97} 
S.~Fan, P.R. Villeneuve, J.D. Joannopoulos, and E.F. Schubert, 
Phys.~Rev.~Lett. $\mathbf{78}$, 3294 (1997).

\bibitem{labilloy97} 
D.~Labilloy, H.~Benisty, C.~Weisbuch, T.F. Krauss, R.M. De~La~Rue, 
V.~Bardinal, R.~Houdr\'e, U.~Oesterle, D.~Cassagne, and C.~Jouanin, 
Phys.~Rev.~Lett. $\mathbf{79}$, 4147 (1997).

\bibitem{fujita98} T.~Fujita, Y.~Sato, T.~Kuitani, and T.~Ishihara,
Phys.~Rev.~B $\mathbf{57}$, 12428 (1998).

\bibitem{benisty99} 
H.~Benisty, C.~Weisbuch, D.~Labilloy, M.~Rattier, C.J.M. Smith, T.F. Krauss, 
R.M. De~La~Rue, R.~Houdr\'e, U.~Oesterle, C.~Jouanin, and D.~Cassagne, 
J.~Lightwave~Technol. $\mathbf{17}$, 2063 (1999).

\bibitem{johnson99} 
S.G. Johnson, S.~Fan, P.R. Villeneuve, J.D. Joannopoulos, 
and L.A.~Kolodziejski, Phys.~Rev.~B $\mathbf{60}$, 5751 (1999).

\bibitem{astratov99}
V.N. Astratov, D.M. Whittaker, I.S. Culshaw, R.M. Stevenson, 
M.S. Skolnick, T.F. Krauss, and R.M. De~La~Rue, 
Phys.~Rev.~B $\mathbf{60}$, R16255 (1999).

\bibitem{whittaker99}
D.M. Whittaker and I.S. Culshaw, 
Phys. Rev. B $\mathbf{60}$, 2610 (1999).

\bibitem{benisty00}
H.~Benisty, D.~Labilloy, C.~Weisbuch, C.J.M. Smith, T.F. Krauss, 
D.~Cassagne, A.~B\'eraud, and C.~Jouanin, 
Appl.~Phys.~Lett. $\mathbf{76}$, 532 (2000).

\bibitem{sondergaard00}
T.~S\o ndergaard, A.~Bjarklev, M.~Kristensen, J.~Erland, and J.~Broeng,
Appl.~Phys.~Lett. $\mathbf{77}$, 785 (2000).

\bibitem{silvestre00}
E.~Silvestre, J.M. Pottage, P.S. Russel, and P.J. Roberts,
Appl.~Phys.~Lett. $\mathbf{77}$, 942 (2000).

\bibitem{loncar00}
M. Lon\v{c}ar, D.~Nedeljkovi\'c, T.~Doll, J.~Vu\v{c}kovi\'c, A.~Scherer, 
and T.~P.~Pearsall, 
Appl.~Phys.~Lett. $\mathbf{77}$, 1937 (2000).

\bibitem{pacradouni00}
V. Pacradouni, W.J. Mandeville, A.R. Cowan, P.~Paddon, J.F. Young, 
and S.R. Johnson, 
Phys.~Rev.~B $\mathbf{62}$, 4204 (2000).

\bibitem{chutinan00}
A. Chutinan and S. Noda,
Phys. Rev. B $\mathbf{62}$, 4488 (2000).

\bibitem{chow00}
E.~Chow, S.Y. Lin, S.G. Johnson, P.R. Villeneuve, J.D. Joannopoulos, 
J.~R. Wendt, G.A. Vawter, W.~Zubrzycki, H.~Hou, and A.~Alleman, 
Nature $\mathbf{407}$, 983 (2000).

\bibitem{ctyroky01} 
J.~\v{C}tyrok\'y, J.~Opt.~Soc.~Am.~A $\mathbf{18}$, 435 (2001).

\bibitem{lalanne01}
P.~Lalanne and H.~Benisty,
J.~Appl.~Phys. $\mathbf{89}$, 1512 (2001).

\bibitem{palamaru01} M.~Palamaru and P.~Lalanne,
Appl.~Phys.~Lett. $\mathbf{78}$, 1466 (2001).

\bibitem{cowan01} A.R. Cowan, P.~Paddon, V.~Pacradouni,
and J.~Young, J.~Opt.~Soc.~Am.~A {\bf 18}, 1160 (2001).

\bibitem{kawai01}
N.~Kawai, K.~Inoue, N.~Carlsson, N.~Ikeda, Y.~Sugimoto, K.~Asakawa, 
and T.~Takemori, 
Phys.~Rev.~Lett. $\mathbf{86}$, 2289 (2001).

\bibitem{ochiai01a}
T. Ochiai and K. Sakoda, Phys.~Rev.~B $\mathbf{63}$, 125107 (2001).

\bibitem{ochiai01b} 
T. Ochiai and K. Sakoda, 
Phys.~Rev.~B $\mathbf{64}$, 045108 (2001).

\bibitem{sakoda_book} 
K.~Sakoda, {\em Optical Properties of
Photonic Crystals} (Springer, Berlin, 2001).

\bibitem{whittaker02} D.M. Whittaker, I.S. Culshaw, V.N. Astratov, 
and M.S. Skolnick, Phys.~Rev.~B $\mathbf{65}$, 073102 (2002).

\bibitem{loncar02}
M.~Lon\v{c}ar, D.~Nedeljkovi\'c, T.P. Pearsall, J.~Vu\v{c}kovi\'c, A.~Scherer,
S.~Kuchinsky, and D.C. Allan, Appl.~Phys.~Lett. $\mathbf{80}$, 1689 (2002).

\bibitem{hadley02}
G.R.~Hadley, IEEE~Photon.~Technol.~Lett. $\mathbf{14}$, 642 (2002).

\bibitem{fan02}
S.~Fan and J.D. Joannopoulos, Phys.~Rev.~B $\mathbf{65}$, 235112 (2002).

\bibitem{tikhodeev02} S.G. Tikhodeev, A.L. Yablonskii, E.A. Muljarov,
N.A. Gippius, and T.~Ishihara, Phys.~Rev.~B $\mathbf{66}$, 045102 (2002).

\bibitem{galli02} M.~Galli, M.~Agio, L.C.~Andreani, L.~Atzeni, D.~Bajoni,
G.~Guizzetti, L.~Businaro, E. Di~Fabrizio, F.~Romanato, and A.~Passaseo,
Eur.~Phys.~J.~B $\mathbf{27}$, 79 (2002).

\bibitem{qiu02} M.~Qiu, Phys. Rev. B $\mathbf{66}$, 033103 (2002).

\bibitem{peyrade02_apl} D.~Peyrade, E.~Silberstein, P.~Lalanne,
A.~Talneau, and Y.~Chen, Appl.~Phys.~Lett. {\bf 81}, 829 (2002).

\bibitem{peyrade02_mne} D.~Peyrade, Y.~Chen, A.~Talneau, M.~Patrini,
M.~Galli, F.~Marabelli, M.~Agio, L.C.~Andreani, E.~Silberstein,
and P.~Lalanne, Microelectron.~Engin. $\mathbf{61-62}$, 529 (2002).

\bibitem{bristow02} A.D. Bristow, V.N. Astratov, R.~Shimada, 
I.S. Culshaw, M.S. Skolnick, D.M. Whittaker, A.~Tahraoui,
and T.F. Krauss, IEEE~J.~Quantum~Electron. $\mathbf{38}$, 880 (2002).

\bibitem{patrini02} M.~Patrini, M.~Galli, F.~Marabelli, M.~Agio,
L.C.~Andreani, D.~Peyrade, and Y.~Chen,
IEEE J.~Quantum~Electron. $\mathbf{38}$, 885 (2002).

\bibitem{andreani02_ieee} L.C. Andreani and M. Agio,
IEEE J.~Quantum~Electron. $\mathbf{38}$, 891 (2002).

\bibitem{andreani02_pss}
L.C. Andreani, Physica Status Solidi (b) $\mathbf{234}$, 139 (2002).

\bibitem{andreani03_apl}
L.C. Andreani and M. Agio, Appl. Phys. Lett. $\mathbf{82}$, 
2011 (2003).

\bibitem{koshino03}
K.~Koshino, Phys. Rev. B {\bf 67}, 165213 (2003).

\bibitem{bristow03}
A.D. Bristow, D.M. Whittaker, V.N. Astratov, M.S. Skolnick,
A.~Tahraoui, T.F. Krauss, M.~Hopkinson, M.P. Croucher, 
and G.A. Gehring, Phys. Rev. B {\bf 68}, 033303 (2003).

\bibitem{yariv}
A. Yariv and P. Yeh, \textit{Optical Waves in Crystals} 
(Wiley, New York, 1984).

\bibitem{joannopoulos_book} 
J.D. Joannopoulos, R.D. Meade, and J.N. Winn,
{\em Photonic Crystals: Molding the Flow of Light} 
(Princeton University Press, Princeton, 1995).


\bibitem{lambda4} For $\epsilon_{\mathrm{diel}}=12$
the $\lambda/4$ condition occurs at $f_{\mathrm{air}}=0.776$:
this corresponds to the first gap being maximum
and to the vanishing of the second-order gap
together with all gaps of even order.



\bibitem{magnusson92}
R. Magnusson and S.S. Wang, 
Appl. Phys. Lett. $\mathbf{61}$, 1022 (1992).



\bibitem{li92}
L.~Li and J.J.~Burke, Opt. Lett. {\bf 17}, 1195 (1992).

\bibitem{popov93} 
E.~Popov, Progr. Optics {\bf 31}, 141 (1993).

\bibitem{neviere95}
M.~Nevi\`ere, E.~Popov and R.~Reinisch,
J. Opt. Soc. Am. A {\bf 12}, 513 (1995).

\bibitem{stancil96}
D.D.~Stancil, Appl. Opt. {\bf 35}, 4767 (1996).

\bibitem{popov_book}
E.G.~Loewen and E.~Popov,
{\em Diffraction Gratings and Applications}
(Dekker, New York, 1997).

\bibitem{tamir97}
T.~Tamir and S.~Zhang, 
J. Opt. Soc. Am. A {\bf 14}, 1607 (1997).

\bibitem{rosenblatt97}
D.~Rosenblatt, A.~Sharon, and A.A. Friesem, 
IEEE J.~Quantum~Electron. {\bf 33}, 2038 (1997). 

\bibitem{norton97}
S.M. Norton, T. Erdogan, and G.M. Morris, 
J. Opt. Soc. Am. A {\bf 14}, 629 (1997).

\bibitem{norton98}
S.M. Norton, G.M. Morris, and T. Erdogan, 
J. Opt. Soc. Am. A {\bf 15}, 464 (1998).

\bibitem{brundrett98}
D.L. Brundrett, E.N. Glytsis, and T.K. Gaylord,
Opt. Lett. {\bf 23}, 700 (1998).

\bibitem{brundrett00}
D.L. Brundrett, E.N. Glytsis, T.K. Gaylord, and J.M. Bendickson,
J. Opt. Soc. Am. A {\bf 17}, 1221 (2000).

\bibitem{neviere_book}
M.~Nevi\`ere, E.~Popov, R.~Reinisch, and G.~Vitrant,
{\em Electromagnetic Resonances in Nonlinear Optics}
(Gordon and Breach, Amsterdam, 2000).

\bibitem{lalanne00}
P.~Lalanne and E.~Silberstein,
Opt. Lett. {\bf 25}, 1092 (2000).

\bibitem{nilsen01}
S.~Nilsen-Hofseth and V.~Romero-Roch\'{\i}n,
Phys. Rev. E {\bf 64}, 36614 (2001).

\bibitem{cao02}
Q.~Cao, P.~Lalanne, and J.-P. Hugonin,
J. Opt. Soc. Am. A {\bf 19}, 335 (2002).



\bibitem{comparison}
The perturbative treatment of the coupling to radiative modes
is analogous to that introduced 
by Ochiai and Sakoda \cite{ochiai01b},
however in the present method the dielectric modulation
described by the tensor $\epsilon_{\mathbf{G},\mathbf{G}'}$
is treated exactly, thereby going beyond the
nearly-free-photon approximation of Ref.~\onlinecite{ochiai01b}.

\bibitem{carniglia71}
C.K.~Carniglia and L.~Mandel, Phys. Rev. D $\mathbf{3}$, 280 (1971).

\bibitem{agranovich85}
V.M. Agranovich, and V.E. Kravtsov, 
Solid State Commun. {\bf 55}, 85 (1985).

\bibitem{tmpol}
It is known in the grating literature that 
the application of coupled-wave analysis 
to the case of TM polarization is more difficult 
and special methods are needed to stabilize numerical 
convergence, as discussed by Ph.~Lalanne and G.~M. Morris,
J. Opt. Soc. Am. A {\bf 13}, 779 (1996).


\bibitem{ho90}
K.M. Ho, C.T. Chan, and C.M. Soukoulis,
Phys. Rev. Lett. $\mathbf{65}$, 3152 (1990).

\bibitem{brillouin}
In Ref.~\onlinecite{brundrett00} a Brillouin diagram
for the losses is shown, i.~e., the imaginary part
of the wavevector is displayed as a function of frequency.
The meaning of the loss diagram is therefore different 
from those in Fig.~\ref{loss}, in particular the imaginary
part of the wavevector is largest in the photonic gap regions.


\end{thebibliography}
\end{document}